\documentstyle[preprint,aps]{revtex}

\begin{document}

\draft

\tighten

\title{Random Matrix Elements and Eigenfunctions in Chaotic Systems}

\author{
Sanjay Hortikar\footnote{E--mail: \tt horti@physics.ucsb.edu}
   and 
Mark Srednicki\footnote{E--mail: \tt mark@tpau.physics.ucsb.edu}
       }

\address{Department of Physics, University of California,
         Santa Barbara, CA 93106 
         \\ \vskip0.5in}

\maketitle

\begin{abstract}
\normalsize{
The expected root-mean-square value of a matrix element $A_{\alpha\beta}$
in a classically chaotic system, where $A$ is a smooth, $\hbar$-independent
function of the coordinates and momenta, and $\alpha$ and $\beta$ label
different energy eigenstates, has been evaluated in the literature in
two different ways: by treating the energy eigenfunctions as gaussian
random variables and averaging $|A_{\alpha\beta}|^2$ over them;
and by relating $|A_{\alpha\beta}|^2$ 
to the classical time-correlation function of $A$.
We show that these two methods give the same answer only if
Berry's formula for the spatial correlations in the energy
eigenfunctions (which is based on a microcanonical density
in phase space) is modified at large separations in a manner
which we previously proposed.
}
\end{abstract}

\pacs{}

Hamiltonian systems which are classically chaotic have quantum
energy eigenvalues, eigenfunctions, and transition matrix elements
which can be profitably analyzed statistically \cite{leshouces,casati}.
Our focus in this paper will be on matrix elements 
(in the energy-eigenstate basis) of operators whose Weyl symbols
are smooth, $\hbar$-independent functions of the classical
coordinates and momenta.  Two different methods have been
proposed in the literature for calculating
the root-mean-square statistical average of these matrix elements
in the limit of small $\hbar$.
One method is to compute this average by treating the energy eigenfunctions
as gaussian random variables; 
the other relates the average to the operator's classical power spectrum.
Our goal is to see whether or not these two methods give the
same result, a question which was first raised 
by Austin and Wilkinson \cite{aw}.
We find that the methods do agree, but only if our recently proposed
modification of Berry's formula \cite{berry77} for the spatial correlations
in energy eigenfunctions of chaotic systems is invoked when
the spatial separation is large compared to any relevant classical
distance scales in the problem \cite{us}.

We begin by reviewing the power-spectrum method, essentially
following the original arguments of Feingold and Peres \cite{fp};
more rigorous treatments leading to the same result have been given 
by Wilkinson \cite{wilk} and Prosen \cite{pros}.
To simplify the discussion, we will consider hermitian operators which
are functions of only the coordinates $\bf q$ (and not the
momenta $\bf p$).  Given a suitable operator $A$ of this type,
we begin by defining
\begin{equation}
F \equiv \int^{+\infty}_{-\infty} dt \,
                        e^{-t^2\!/2\tau_{\rm c}^2}
                        e^{i\omega t}
                        \langle\alpha|A_t A|\alpha\rangle \;,
\label{f}
\end{equation}
where $A_t\equiv e^{iHt/\hbar}A e^{-iHt/\hbar}$ 
is the relevant operator at time $t$ in the Heisenberg-picture, 
$|\alpha\rangle$ is an energy eigenstate with energy $E_\alpha$,
$\omega$ is a parameter, and $\tau_{\rm c}$ is a time cutoff
which may be needed for convergence of the integral.

We now evaluate $F$ in two different ways.
First, we use Shnirelman's theorem \cite{shnir,zeld,cdv,hmr,zz},
which says that, in the limit of small $\hbar$, 
the expectation value of an operator $O$ in an energy eigenstate 
is equal to its classical, microcanonical average at the corresponding energy,
\begin{equation}
\langle\alpha|O|\alpha\rangle = \int d\mu_{E_\alpha}\, O_W({\bf p},{\bf q})\; ,
\label{shni}
\end{equation}
where $O_W({\bf p},{\bf q})$ is the Weyl symbol of the operator $O$, and
$d\mu_E$ denotes the Liouville measure on the surface in phase
space with energy $E$, 
\begin{equation}
d\mu_E = {1\over\bar\rho(E)}\;
         {d^f\! p \, d^f\! q \over (2\pi\hbar)^f}\; 
         \delta(E-H_W({\bf p},{\bf q})) \; .
\label{mue}
\end{equation}
Here $f$ is the number of degrees of freedom,
$H_W({\bf p},{\bf q})$ is the Weyl symbol of the hamiltonian operator $H$,
and $\bar\rho(E)$ is the semiclassical density of states,
\begin{equation}
{\bar\rho}(E) = \int {d^f\! p \, d^f\! q \over (2\pi\hbar)^f}\; 
                     \delta(E-H_W({\bf p},{\bf q})) \; .
\label{rho}
\end{equation}
Note that $d\mu_E$ is a purely classical object; the factors
of $\hbar$ cancel between eqs.~(\ref{mue}) and (\ref{rho}).
Also, Shnirelman's theorem is proved for principal symbols instead
of Weyl symbols, but there is no difference in the $\hbar\to 0$ limit 
which the theorem also requires; see \cite{back} for a thorough discussion.

We now apply eq.~(\ref{shni}) in eq.~(\ref{f}), making the approximation
(valid in the $\hbar\to0$ limit) that 
\begin{equation}
\int d\mu_E \, (A_t A)_W = \int d\mu_E \, A_W({\bf q}_t)A_W({\bf q}) \;,
\label{weyl}
\end{equation}
where $A_W({\bf q})$ is the Weyl symbol of the operator $A$
(which, by assumption, depends only on ${\bf q}$ and not ${\bf p}$),
and ${\bf q}_t$ is the classical coordinate at time $t$, assuming
an initial point $({\bf p},{\bf q})$ on the surface with energy $E$
in phase space.  We therefore obtain
\begin{equation}
F =  \int^{+\infty}_{-\infty} dt \,
                        e^{-t^2\!/2\tau_{\rm c}^2}
                        e^{i\omega t}
                        \int d\mu_{E_\alpha} \, 
                        A_W({\bf q}_t)A_W({\bf q}) \; .
\label{fp1}
\end{equation}

We now evaluate $F$ in a different manner:
we insert a complete set of energy eigenstates to get
\begin{eqnarray}
F &=& \sum_\beta \int^{+\infty}_{-\infty} dt \,
                     e^{-t^2\!/2\tau_{\rm c}^2}
                     e^{i\omega t}
                     \langle\alpha|A_t|\beta\rangle
                     \langle\beta|A|\alpha\rangle
\nonumber \\
                 &=& \sum_\beta \int^{+\infty}_{-\infty} dt \,
                     e^{-t^2\!/2\tau_{\rm c}^2}
                     e^{i(E_\alpha-E_\beta+\hbar\omega)t/\hbar}
                     A_{\alpha\beta} A_{\beta\alpha}
\nonumber \\
                 &=& 2\pi\hbar\sum_\beta 
                     \delta_{\hbar/\tau_{\rm c}}(E_\alpha-E_\beta+\hbar\omega)
                     |A_{\alpha\beta}|^2 \;,
\label{fp2}
\end{eqnarray}
where $A_{\alpha\beta}\equiv\langle\alpha|A|\beta\rangle$, and 
$\delta_\varepsilon(E)$ denotes a smeared delta function with a width of
$\varepsilon$.
We now assume that each eigenstate has a random character, so that,
with $E_\alpha$ and $E_\beta$ each varied over a small range,
there is a smooth distribution of values for $|A_{\alpha\beta}|^2$.
Let this distribution be characterized by an expected value 
which we will call $\langle |A_{\alpha\beta}|^2\rangle$.
If we also take the width $\hbar/\tau_{\rm c}$ of the smeared delta functions
to be somewhat larger than the mean level spacing,
equal to $1/\bar\rho(E)$ in the limit of small $\hbar$,
we can replace the sum over $\beta$ in eq.~(\ref{fp2}) by an integral over 
$\bar\rho(E_\beta) dE_\beta$.  Thus we have
\begin{eqnarray}
F &=& 2\pi\hbar \int_0^\infty dE_\beta \,\bar\rho(E_\beta) 
                      \delta_{1/\bar\rho}(E_\alpha-E_\beta+\hbar\omega)
                      \langle|A_{\alpha\beta}|^2\rangle
\nonumber \\
  &=& 2\pi\hbar \, \bar\rho(E_\alpha+\hbar\omega) \,
      \langle|A_{\alpha\beta}|^2\rangle \; .
\label{fp3}
\end{eqnarray}
Equating the right-hands sides of eqs.~(\ref{f}) and (\ref{fp3})
gives us the desired formula for $\langle|A_{\alpha\beta}|^2\rangle$;
however, its accuracy to subleading order in $\hbar$ can be improved 
by symmetrizing on $E_\alpha$ and $E_\beta$ \cite{pros} to get
\begin{equation}
\langle|A_{\alpha\beta}|^2\rangle 
= {1\over\tau_{\rm H}}
  \int^{+\infty}_{-\infty} dt \, 
                           e^{-2\pi^2 t^2\!/\tau^2_{\rm H}}
                           e^{i\omega t}
                           \int d\mu_{\bar E} \, 
                           A_W({\bf q}_t)A_W({\bf q}) \; ,
\label{aab0}
\end{equation}
where $\bar E = {1\over2}(E_\alpha+E_\beta)$ is the mean energy,
$\hbar\omega = E_\beta-E_\alpha$ is the energy difference, and
$\tau_{\rm H} = 2\pi\hbar\bar\rho(\bar E)$ is the Heisenberg time.
If we hold $\bar E$ and $\omega$ fixed in the limit of small $\hbar$,
the right-hand side of eq.~(\ref{aab0}) is simply $1/\tau_{\rm H}$
times the classical power spectrum of the observable $A$ at energy $\bar E$,
with any structure on frequency scales less than $2\pi/\tau_{\rm H}$
smeared out by the time cutoff.
Eq.~(\ref{aab0}) is the first of two formulas for 
$\langle|A_{\alpha\beta}|^2\rangle$ which can be found in the literature.

We get the second formula \cite{sred94,eck2,ss96}
by first writing the squared matrix element
in terms of the eigenfunctions as
\begin{equation}
|A_{\alpha\beta}|^2
= \int d^f\! q' \, \psi_\alpha^*({\bf q'}) A_W({\bf q}') \psi_\beta({\bf q}') 
  \int d^f\! q  \, \psi_\beta^*({\bf q})   A_W({\bf q})  \psi_\alpha({\bf q}) 
       \; .
\label{aab1}
\end{equation}
In a chaotic system, the individual eigenfunctions can be treated
as independent random variables with a gaussian probability 
distribution \cite{berry77,ogh,pei,paei,prig1,prig2,alhas,sred96}.
Because it is gaussian, this
distribution is completely specified by the two-point correlation function
\begin{equation}
C({\bf q}',{\bf q}|E) 
\equiv \langle \psi({\bf q}')\psi^*({\bf q}) \rangle \; ,
\label{cexp}
\end{equation}
where the angle brackets denote averaging over the probability distribution
for $\psi({\bf q})$ given the energy $E$.
Averaging eq.~(\ref{aab1}) over this probability distribution yields
\begin{equation}
\langle|A_{\alpha\beta}|^2\rangle 
= \int d^f\! q' \, d^f\! q \; C({\bf q},{\bf q}'|E_\alpha) 
                              A_W({\bf q}')
                              C({\bf q}',{\bf q}|E_\beta) 
                              A_W({\bf q}) \; .
\label{rand}
\end{equation}
This is the second formula for $\langle|A_{\alpha\beta}|^2\rangle$
which can be found in the literature.
The question is whether or not it is the same as the first formula,
eq.~(\ref{aab0}).

Of course, in order to answer this question we need an explicit
expression for $C({\bf q},{\bf q}'|E)$.
Berry \cite{berry77} conjectured that, in the small-$\hbar$ limit,
\begin{equation}
C({\bf q}',{\bf q}|E) 
= \int d\mu_E\, e^{i{\bf P}\cdot({\bf q}'-{\bf q})/\hbar}
    \delta({\bf Q}-{\textstyle{1\over2}}({\bf q}'+{\bf q})) \; ,
\label{berry}
\end{equation}
where the Liouville integral is over $({\bf P},{\bf Q})$.
However \cite{aw}, this formula for $C({\bf q}',{\bf q}|E)$
appears to be too simple to be able reproduce the classical power spectrum of
$A$ which appears in eq.~(\ref{aab0}).

In a separate paper \cite{us}, we have argued that, in fact, 
eq.~(\ref{berry}) must be modified whenever the separation 
$|{\bf q}'-{\bf q}|$ is large, in the sense that the shortest classical path 
with energy $E$ which connects $\bf q$ to ${\bf q}'$ is not well approximated 
by a linear function of time.  This will be generically true in
eq.~(\ref{aab0}), since both $\bf q$ and ${\bf q}'$ are integrated,
and since the factors of $A_W({\bf q})$ and $A_W({\bf q}')$ do not
force $\bf q$ and ${\bf q}'$ to be close together.
When $\bf q$ and ${\bf q}'$ are far apart,
eq.~(\ref{berry}) should be replaced with
\begin{equation}
C({\bf q}',{\bf q}|E) = {2\over{\bar\rho}(E)(2\pi\hbar)^{(f+1)/2}}
                           \sum_{\rm paths} |D_{\rm p}|^{1/2}
                           \cos[S_{\rm p}/\hbar-(2\nu_{\rm p}+f-1)\pi/4] \; ,
\label{cnew}
\end{equation}
where the sum is over all classical paths connecting 
${\bf q}$ to ${\bf q}'$ with energy $E$; each path has
action $S_{\rm p} = \int_{\bf q}^{{\bf q}'} {\bf p} \!\cdot\! d{\bf q}$,
focal point number $\nu_{\rm p}$, and fluctuation determinant 
\begin{equation}
D_{\rm p} = \det\pmatrix{ {\partial^2 S_{\rm p} \over
                           \partial{\bf q} \partial{\bf q}' } &
                          {\partial^2 S_{\rm p} \over
                           \partial E \partial{\bf q}'        } \cr
\noalign{\medskip}
                          {\partial^2 S_{\rm p} \over
                           \partial{\bf q} \partial E         } &
                          {\partial^2 S_{\rm p} \over
                           \partial E^2                       } \cr } .
\label{det1}
\end{equation}
Eq.~(\ref{cnew}) actually holds only if the system is invariant
under time reversal; otherwise a more cumbersome formula is needed \cite{us}.  
The final formula for $\langle|A_{\alpha\beta}|^2\rangle$
turns out to be the same in either case, 
and so to simplify the notation we will use eq.~(\ref{cnew}).
Eq.~(\ref{cnew}), or its replacement for a system which is not
time-reversal invariant, is valid as long as the contributing
path of least action has $S_{\rm p}/\hbar \gg 1$.  
This is of course true generically in the limit of small $\hbar$.

We now show that if we use eq.~(\ref{cnew}) for $C({\bf q}',{\bf q}|E)$,
eq.~(\ref{rand}) gives the same result for 
$\langle|A_{\alpha\beta}|^2\rangle$ as eq.~(\ref{aab0}).

We begin by substituting eq.~(\ref{cnew}) into eq.~(\ref{rand}).
Since we are interested in the limit of small $\hbar$ with 
$\bar E$ and $\omega$ held fixed, we can usually replace $E_\alpha$
and $E_\beta$ with $\bar E$.  
We then make use of the ``diagonal approximation'' \cite{berry85}
in which the double sum over paths is collapsed to a single sum.
In related calculations \cite{berry85,eck2},
this can be justified by the rapidly oscillating phases of the 
off-diagonal terms as long as the single sum includes only
those paths whose elapsed times are less than the Heisenberg time. 
We assume the same condition holds here.
The product of cosines in each diagonal term then yields
a single cosine which is slowly oscillating, and we get
\begin{equation}
\langle|A_{\alpha\beta}|^2\rangle
= {2\over{\bar\rho}(\bar E)^2(2\pi\hbar)^{f+1}}
  \int d^f\!q'\,d^f\!q\,A_W({\bf q}')A_W({\bf q})
  \sum_{\rm paths}
  |D_{\rm p}| \cos(\omega\tau_{\rm p}) \; .
\label{aab3}
\end{equation}
The sum is over paths from ${\bf q}$ to ${\bf q}'$ with energy $\bar E$,
and elapsed time
\begin{equation}
\tau_{\rm p} = {\partial S_{\rm p}\over\partial E}\Biggr|_{E=\bar E} 
\label{tau}
\end{equation}
less than the Heisenberg time $\tau_{\rm H}$.
We have implicitly assumed that the topological quantity $\nu_{\rm p}$
does not change as the energy of the path is varied from 
$\bar E - {1\over2}\hbar\omega$ to $\bar E + {1\over2}\hbar\omega$.

To make further progress we need to rewrite the fluctuation determinant as
\begin{equation}
D_{\rm p} =   \det\pmatrix{ -{\partial{\bf p}  \over
                              \partial{\bf q}'                   } &
                             {\partial\tau     \over
                              \partial{\bf q}'                   } \cr
\noalign{\medskip}
                            -{\partial{\bf p}  \over
                              \partial E                         } &
                             {\partial\tau     \over
                              \partial E                         } \cr } \; .
\label{det2}
\end{equation}
Here ${\bf p} = -\partial S_{\rm p}/\partial{\bf q}$
is the momentum at the beginning of the path, 
and $\tau=\tau_{\rm p}$ is the elapsed time along the path,
given by eq.~(\ref{tau}).  With these definitions,
eq.~(\ref{det2}) follows immediately from eq.~(\ref{det1}).
Eq.~(\ref{det2}) shows us that $|D_{\rm p}|$ can be thought of as a 
jacobian for a change of variables from the final position ${\bf q}'$
and total energy $\bar E$ to the initial momentum ${\bf p}$ and elapsed time
$\tau$ \cite{gutz1}.  To make use of this, we insert
$1=\int d\bar E\,\delta(\bar E - H_W({\bf p},{\bf q}))$ 
into eq.~(\ref{aab3}) and change variables.  We now have
\begin{equation}
\langle|A_{\alpha\beta}|^2\rangle
= {1\over\pi\hbar{\bar\rho}(\bar E)^2}
  \int_0^{\tau_{\rm H}} d\tau
  \int {d^f\! p\, d^f\! q \over (2\pi\hbar)^f} 
  \sum_{\rm paths}
  \delta(\bar E - H_W({\bf p},{\bf q}))
  \cos(\omega\tau) 
  A_W({\bf q}')A_W({\bf q}) \; .
\label{aab4}
\end{equation}
The sum is over all paths which begin at $({\bf p},{\bf q})$ and have
elapsed time $\tau$.  However, there is only one such path, 
and so the sum over paths may be dropped.
Also, ${\bf q}'$ is the position at time $\tau$, and it is now more properly
denoted ${\bf q}_\tau$.  Using eq.~(\ref{mue}) and 
$\tau_{\rm H} = 2\pi\hbar\bar\rho(\bar E)$,
we see that eq.~(\ref{aab4}) can be rewritten as
\begin{equation}
\langle|A_{\alpha\beta}|^2\rangle
=   {2\over\tau_{\rm H}}
    \int_0^{\tau_{\rm H}} d\tau
    \cos(\omega\tau) 
    \int d\mu_{\bar E}\,
    A_W({\bf q}_\tau)A_W({\bf q}) \; .
\label{aab6}
\end{equation}
Using the fact that time-translation invariance implies that 
$\int d\mu_{\bar E}\, A_W({\bf q}_\tau)A_W({\bf q})$ is an even function
of $\tau$ (even if the system is not time-reversal invariant), 
we see immediately that eq.~(\ref{aab6}) is equivalent to eq.~(\ref{aab0}),
up to the issue of the detailed treatment of the large-time cutoff.
This is our main result.

Another quantity of interest is the size of the fluctuations in the
diagonal matrix elements $A_{\alpha\alpha}$.
If we first shift $A$ (if necessary) so that
$\langle A_{\alpha\alpha}\rangle = \int d\mu_{E_\alpha} \, A_W({\bf q}) = 0$,
then the object we wish to evaluate is $\langle |A_{\alpha\alpha}|^2\rangle$.
This has been done previously \cite{wilk2,eck1,eck2}
by making use of the trace formula \cite{gutz1,gutz2,eck3}
and properties of periodic orbits.
Here we will compute $\langle |A_{\alpha\alpha}|^2\rangle$
by averaging over the gaussian probability distribution for 
energy eigenfunctions \cite{sred94,eck2,ss96}.
In the case of a system which is not invariant 
under time reversal, the energy eigenfunctions are generically complex, 
and the relevant formula is \cite{sred96}
\begin{equation}
\langle\psi^*_1\psi_2\psi^*_3\psi_4\rangle
=
\langle\psi^*_1\psi_2\rangle
\langle\psi^*_3\psi_4\rangle
+
\langle\psi^*_1\psi_4\rangle
\langle\psi^*_3\psi_2\rangle
\; ,
\label{wick1}
\end{equation}
where $\psi_i=\psi_\alpha({\bf q}_i)$.
If the system is invariant under time reversal,
the energy eigenfunctions are real, and we have instead \cite{sred96}
\begin{equation}
\langle\psi_1\psi_2\psi_3\psi_4\rangle
=
\langle\psi_1\psi_2\rangle
\langle\psi_3\psi_4\rangle
+
\langle\psi_1\psi_4\rangle
\langle\psi_2\psi_3\rangle
+
\langle\psi_1\psi_3\rangle
\langle\psi_2\psi_4\rangle
\; .
\label{wick2}
\end{equation}
Combined with the previous results for $\langle|A_{\alpha\beta}|^2\rangle$,
we find that
\begin{equation}
\langle |A_{\alpha\alpha}|^2\rangle
= {g \over \tau_{\rm H}}
  \int^{\tau_{\rm H}}_{-\tau_{\rm H}} d\tau
                        \int d\mu_{E_\alpha} \, 
                        A_W({\bf q}_t)A_W({\bf q}) \; .
\label{aaa}
\end{equation}
Here $g=2$ for a system which is invariant under time reversal,
and $g=1$ for a system which is not.
Eq.~(\ref{aaa}) is in agreement with the results of \cite{wilk2,eck1,eck2}.

\begin{acknowledgments}

We thank Michael Wilkinson for bringing the issue treated in this paper
to our attention.  
This work was supported in part by NSF Grant PHY--97--22022.

\end{acknowledgments}

\end{document}